\documentclass[12pt]{iopart}
\usepackage{iopams}
\usepackage[dvips]{graphicx}
\usepackage{verbatim}
\usepackage{bm}

\begin{document}

\title[]{Coherent population trapping in a dressed two-level atom via a bichromatic field}
\author{Peng Li$^{1,2}$, Xi-Jing Ning$^3$, Qun-Zhang$^2$ and J Q You$^1$}
\address{$^1$Department of Physics and
Surface Physics Laboratory (National Key Laboratory), Fudan
University, Shanghai 200433, China}
\address{$^2$Department of Materials Science, Fudan University, Shanghai 200433, China}
\address{$^3$Institute of Modern Physics and the Key Lab of Applied
Ion Beam Physics of the Ministry of Education, Fudan University,
Shanghai 200433, China}
\eads{\mailto{lipeng@fudan.edu.cn},\mailto{jqyou@fudan.edu.cn}}

\date{\today}

\begin{abstract}
We show theoretically that by applying a bichromatic electromagnetic
field, the dressed states of a monochromatically driven two-level
atom can be pumped into a coherent superposition termed as
dressed-state coherent population trapping. Such effect can be
viewed as a new doorknob to manipulate a two-level system via its
control over dressed-state populations. Application of this effect
in the precision measurement of Rabi frequency, the unexpected
population inversion and lasing without inversion are discussed to
demonstrate such controllability.
\end{abstract}

\pacs{42.50.Hz, 32.80.Qk} \maketitle

%%%%%%%%%%%%%%%%%%%%%%%%%%%%%%%%%%%%%%%%%%%%%%%%%%%%%%%%%%%%%%%%%%%
%%%%%%%%%%%%%%%%%%%%%%%%%%%%%%%%%%%%%%%%%%%%%%%%%%%%%%%%%%%%%%%%%%%
\section{Introduction}\label{sec:1}

As unveiled by the remarkable suppression of fluorescence, coherent
population trapping (CPT)~\cite{Arimondo} has been regarded as a
significant demonstration of macroscopic quantum coherence and
interference. The importance of CPT is well illuminated by its
numerous applications in laser cooling, stimulated Raman adiabatic
passage (STIRAP)~\cite{STIRAP}, electromagnetically induced
transparency (EIT)~\cite{EIT}, and lasing without inversion
(LWI)~\cite{LWI}. Recent studies even extend the concept of CPT
beyond optics to the realm of electron transportation in coupled
quantum dots, where an all-electronic analogy of CPT emerges with
potential values for current rectification~\cite{QDCPT}.

In general, CPT is realized within a driven $\Lambda$-type atom, in
which a \textit {dark state} is prepared via the field-induced
coherent superposition of the two lower atomic states. The mixing
ratio of this superposition is self-adjusted to have the two
transition paths from the lower states to the upper state interfere
destructively, so that excitation to the upper state is forbidden.
Therefore, suppression of fluorescence from the upper state results
in a striking darkline at the vicinity of two-photon
resonance~\cite{Gray}. Since CPT by its nature relies on the
coherence between the two lower atomic states, it is vulnerable to
any mechanism which may destroy this coherence. For this reason, CPT
is predominantly studied in bare $\Lambda$-type atoms with
negligible spontaneous transition between the two lower states.

In many cases the dressed-atom picture provides a unique insight in
dealing with atom-photon interaction problems~\cite{Allen&Eberly,
Scully}. A well-known example is the Rabi splitting of a
monochromatically driven two-level atom (TLA), where the Rabi
doublet can be well treated as eigenstates of the (semiclasical)
atom-field interaction Hamiltonian, i.e., \textit {dressed states}
(DSs). When driving field is modeled quantum-mechanically, a ladder
of such dressed-state doublets (DSDs) emerges with quasi-equidistant
energy spacing between adjacent DSDs. Coherent transition between
DSs within each DSD is forbidden due to vanishing dipole
moment~\cite{Tannoudji}. To induce coherence in the DS basis one
needs to couple adjacent DSDs, where transitions are allowed between
DSs associated with different DSDs~\cite{Tannoudji}. Such coherent
transitions between DSs result in an interesting spectral signature
of the underlying atom-field interation, and novel effects such as
spectral cancellation of spontaneous emission can be realized (see,
the \textit {doubly dressed-atom} in~\cite{Ficek}).

In contrast to~\cite{Ficek},  we show in this paper that it is also
possible to achieve a CPT-like coherence \textit {within} each DSD,
via a bichromatic field which couples the two DSs to an auxiliary
atomic state. We show that such a dressed-state coherent population
trapping (DSCPT) is formed in analogy to the conventional CPT in
bare atomic basis. As expected, the occurence of such DSCPT is
reflected in a sharp suppression of fluorescence from the auxiliary
state. Our analytical and numerical studies show that the efficiency
of DSCPT is limited by its rate of decoherence, i.e., the rate of
spontaneous decay of the TLA (see Section~\ref{sec:2} for details).
Practically, DSCPT can be viewed as a new doorknob to manipulate a
TLA due to its ability to control the DS populations. Application of
DSCPT in the Autler-Townes spectroscopy leads to an enhanced
precision in the measurement of Rabi frequency. DSCPT can also be
used to obtain an unexpected population inversion for a driven TLA,
where the previously reported \textit{dynamically induced
irreversibility}~\cite{meduri} can be treated as a limiting case
within this context, which provides us a new picture in
understanding the origin of this phenomenon. Additionally, novel LWI
effects with and without hidden inversion are also presented to
demonstrate the versatile controllability provided by DSCPT (see
Section~\ref{sec:3} for details).

We wish to emphasize that DSCPT is different from the CPT
investigated in a degenerate two-level atom (DTLA)~\cite{DTLA},
where effects of Zeeman sublevels are investigated. In a DTLA,
sublevels of the ground (excited) level do not contain any
contamination from the excited (ground) level, with or without
external magnetic field. While in a coherently driven TLA, each DS
is a coherent mixing of the ground state and the excited state.
Therefore, the coherence between DSs discussed here intrinsically
differs from that studied in~\cite{DTLA}.

This paper is organized as follows: In Section~\ref{sec:2}, we
introduce the theoretical model and provide both analytical and
numerical investigation of DSCPT. In Section~\ref{sec:3}, we
demonstrate the application of DSCPT via three examples. Finally, a
brief summary is given in Section~\ref{sec:4}.

%%%%%%%%%%%%%%%%%%%%%%%%%%%%%%%%%%%%%%%%%%%%%%%%%%%%%%%%%%%%%%%%%%%
%%%%%%%%%%%%%%%%%%%%%%%%%%%%%%%%%%%%%%%%%%%%%%%%%%%%%%%%%%%%%%%%%%%
\section{Demonstration of dressed-state coherent population trapping}\label{sec:2}
\subsection{Model and theoretical analysis}
%%%%%%%%%%%%%%%%%%%%%%%%%%%%%%%%%%%%%%%%%%%%%%%%%%%%%%%%%%%%%%%%%%%%
%Figure 1
\begin{figure}[htb]
\begin{center}
\includegraphics[bb=120 558 410 714, width=8cm, clip]{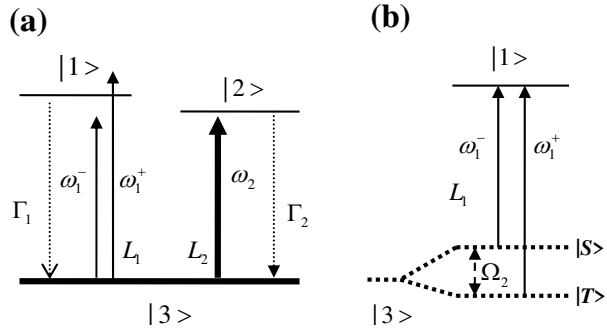}
\end{center}
\caption {Schematic diagram of a V-type three-level atom driven by
two externally applied electromagnetic fields. (a) The $\vert
1\rangle -\vert 3\rangle $ transition of the V-type atom is driven
by a weak bichromatic field $L_1$, while the $\vert 2\rangle -\vert
3\rangle $ transition is driven by a strong monochromatic field
$L_2$. (b) An effective $\Lambda$-type atom formed in (a), where
$\vert S\rangle$ and $\vert T\rangle$ are the DSs of the subsystem
composed of transition $\vert 2\rangle - \vert 3\rangle$ and field
$L_2$.}\label{fig1}
\end{figure}
%%%%%%%%%%%%%%%%%%%%%%%%%%%%%%%%%%%%%%%%%%%%%%%%%%%%%%%%%%%%%%%%%%%%

The theoretical model we study is a V-type atom [see
Fig.~\ref{fig1}(a)]. The $\vert 1\rangle -\vert 3\rangle $
transition is driven by a weak bichromatic field $L_1$, while the
$\vert 2\rangle -\vert 3\rangle $ transition is driven by a strong
monochromatic field $L_2$. The angular frequencies corresponding to
these two transitions are $\omega_{13}$ and $\omega_{23}$,
respectively. The two driving fields $L_1$ and $L_2$ are expressed
as $E_1(t)=\frac{1}{2}E_1^+\cdot {\rm e}^{{\rm
i}\omega_1^+t}+\frac{1}{2}E_1^-\cdot {\rm e}^{{\rm
i}\omega_1^-t}+{\rm H.c}$ and $E_2(t)=\frac{1}{2}E_2\cdot {\rm
e}^{{\rm i}\omega_2t}+{\rm H.c}$, where
$\omega_1^\pm=\omega_1\pm\delta$ denote the two frequencies of the
bichromatic field $L_1$ and $\omega_2$ is the frequency of the
monochromatic field $L_2$. Thus we can define two detunings
$\Delta_1=\omega_{13}-\omega_1$ and $\Delta_2=\omega_{23}-\omega_2$.
The Rabi frequencies associated with the two atomic transitions can
be written as $\Omega_1^\pm=E_1^\pm\cdot \mu_{13}/\hbar$ and
$\Omega_2=E_2\cdot \mu_{23}/\hbar$, where $\mu_{13}$($\mu_{23}$)
denotes the dipole moment of the $\vert 1\rangle -\vert
3\rangle $ ($\vert 2\rangle -\vert 3\rangle $) transition. The rate
of spontaneous decay from $\vert 1 \rangle$ ($\vert 2 \rangle$) to
$\vert 3 \rangle$ is denoted as $\Gamma_1$($\Gamma_2$). In what
follows, the field amplitudes $E_1^\pm$ and $E_2$, the dipole
moments $\mu_{13}$ and $\mu_{23}$, and the Rabi frequencies
$\Omega_1^\pm$ and $\Omega_2$ are chosen as real numbers for
simplicity, without loss of generality.

First let us look at the system as if $L_1$ is absent. It is obvious
that $L_2$ plus $\vert 2 \rangle - \vert 3 \rangle$ transition
correspond to a standard driven TLA~\cite{Mollow}. The semiclassical
DSs are the eigenstates of the Hamiltonian (in the rotating-wave
frame, $\hbar=1$) $H_{\rm TLA}=\Delta_2 \vert \widetilde 2\rangle
\langle \widetilde 2 \vert + \frac{1}{2} \Omega_2(\vert \widetilde 3
\rangle \langle \widetilde 2 \vert+ \vert\widetilde 2\rangle
\langle\widetilde 3 \vert)$, i.e.,
\begin{eqnarray}\label{eq.ds.semi}
\vert S\rangle &=&\sin \chi \vert \widetilde 3\rangle +\cos\chi\vert
\widetilde 2\rangle,\nonumber \\ \vert T\rangle &=&\cos \chi
\vert\widetilde 3\rangle -\sin \chi \vert\widetilde 2\rangle,
\end{eqnarray}%
where $\chi=\frac{1}{2}\tan^{(-1)} (\Omega_2/\Delta_2)$, $ \vert
\widetilde 2\rangle=\vert 2\rangle{\rm e}^{-{\rm i}\omega_2 t}$, and
$\vert \widetilde 3 \rangle=\vert 3\rangle$. And the corresponding
eigenenergies are $E_{S,T}=\frac{1}{2}\Delta_2 \pm
\frac{1}{2}\sqrt{\Delta_2^2 +\Omega_2^2}$ , with $G=\sqrt{\Delta_2^2
+\Omega_2^2}$ being the the Rabi-splitting. (We emphasize that both
the ground state $\vert 3\rangle$ and the upper state $\vert
2\rangle$ are splitted into doublet via the atom-field interaction.
Without specific notation, we refer $\vert S\rangle$ and $\vert
T\rangle$ to the field-induced doublet associated with the ground
state $\vert 3\rangle$, for the rest of this paper.) Meanwhile, the
quantum mechanical DSs are the eigenstates of the Hamiltonian
${{\cal H}'_{\rm TLA}}=\Delta_2\vert 3,n\rangle\langle 3,
n\vert+g\sqrt{n}\left(\vert 3,n\rangle\langle 2, n-1\vert+{\rm
H.c}\right)$, i.e.,
\begin{eqnarray}\label{eq.ds.quant}
\vert S_n\rangle &=&\sin \chi' \vert 3,n\rangle
+\cos\chi'\vert 2,n-1\rangle,\nonumber\\
\vert T_n\rangle &=&\cos \chi' \vert 3,n\rangle -\sin \chi' \vert
2,n-1\rangle,
\end{eqnarray}%
where $\chi'=\frac{1}{2}\tan^{(-1)} (2g\sqrt{n}/\Delta_2)$, with
$\vert n\rangle$ and $g$ being the Fock state of the field and the
coupling strength between the field and the atom, respectively.
Since the classical field $L_2$ is not in a pure Fock state $\vert
n\rangle$ but a coherent state with average photon number $\bar n$,
the correspondence between semiclassical Rabi frequency and the
field's photon number leads to $\Omega_2=2g\sqrt{\bar n}$.

To achieve CPT in the above DS basis one needs to construct an
effective $\Lambda$-type atom. This can be fulfilled by introducing
the bichromatic field $L_1$, whose center frequency is (near)
resonant with $\vert 1\rangle -\vert 3\rangle $ transition, as shown
in Fig.~{\ref{fig1}}(a). The two transitions $\vert 1\rangle -\vert
T\rangle$ and $\vert 1\rangle -\vert S\rangle$ are then coupled by
the two frequency components $\omega_1^+$ and $\omega_1^-$ of $L_1$,
respectively, as shown in Fig.~{\ref{fig1}}(b). Since both $\vert
S\rangle$ and $\vert T\rangle$ are contaminated by bare state $\vert
3 \rangle$, the effective dipole moments corresponding to
transitions $\vert 1\rangle -\vert S\rangle$ and $\vert 1\rangle
-\vert T\rangle$ can easily be deduced from Eq.~(\ref{eq.ds.semi})
as $ \mu_{1S}=\langle 1\vert\vec{\cal D}\vert
S\rangle=\mu_{13}\sin\chi$ and $ \mu_{1T}=\langle 1\vert\vec{\cal
D}\vert T\rangle=\mu_{13}\cos\chi$. Neglecting the non-resonant
terms (which involves a second RWA), i.e., the coupling between the
field component $\omega_1^+$ and $\vert 1\rangle-\vert S\rangle$
transition, and that between $\omega_1^-$ and $\vert 1\rangle-\vert
T\rangle$ transition, the Rabi frequencies associated with these two
transitions are simply $ \Omega_{S}=\Omega_1^-\sin\chi$ and
$\Omega_{T}=\Omega_1^+\cos\chi$. Accordingly, the Hamiltonian of the
effective $\Lambda$-type atom in Fig.~\ref{fig1}(b) can be written
as
\begin{eqnarray}\label{eq.hlambdaprime}
H_{\Lambda}&=& \frac{1}{2}\left[\Omega_{S}\vert\widetilde 1\rangle
\langle \widetilde S\vert + \Omega_{T}\vert\widetilde
1\rangle\langle\widetilde T\vert+{\rm H.c}\right]\nonumber
\\ &+&\Delta\vert\widetilde 1\rangle\langle\widetilde 1\vert+
\Delta_{\rm 2photon}\vert\widetilde S\rangle\langle\widetilde
S\vert,
\end{eqnarray}%
where $\Delta=\omega_{13}-E_T-\omega_1^+$, $\Delta_{\rm
2photon}=(E_S-E_T)-(\omega_1^+-\omega_1^-)$, $ \vert\widetilde
1\rangle={\rm e}^{-{\rm i}\omega_1^+t}\vert 1\rangle$,
$\vert\widetilde S\rangle={\rm e}^{-{\rm
i}(\omega_1^+-\omega_1^-)t}\vert S\rangle$, and $\vert\widetilde
T\rangle=\vert T\rangle$. As essential for CPT, the two-photon
resonance condition requires that
$(E_S-E_T)-(\omega_1^+-\omega_1^-)=0$, i.e.,
\begin{equation}\label{eq.2photon}
G=2\delta,
\end{equation}%
which can fulfilled by manipulating either $L_1$ or $L_2$, via
tuning $2\delta$ or $G$, respectively.

Once Eq.(\ref{eq.2photon}) is satisfied, it is routine to show that
both dark and bright states exist in this effective $\Lambda$-type
atom. We adopt the terminology in~\cite{Arimondo}, i.e., the
noncoupled and coupled states, which are defined as
\begin{eqnarray}\label{eq.NC}
\vert NC \rangle &=& \cos \theta \vert\widetilde S \rangle - \sin
\theta \vert\widetilde T \rangle\nonumber\\
&=&{\rm e}^{-{\rm i}2 \delta t}\cos \theta \vert S \rangle - \sin
\theta \vert T \rangle,
\end{eqnarray}%
and
\begin{eqnarray}\label{eq.C}
\vert C \rangle &=& \sin \theta \vert\widetilde S \rangle
+\cos\theta\vert\widetilde T \rangle\nonumber\\
&=& {\rm e}^{-{\rm i}2 \delta t}\sin \theta \vert S \rangle
+\cos\theta \vert T \rangle,
\end{eqnarray}%
where $\theta=\tan^{-1}(\Omega_S /\Omega_T)$. Since $\langle 1\vert
H'_{\Lambda} \vert NC\rangle =0$, an atom in the noncoupled state
$\vert NC \rangle$ cannot be excited to $\vert 1\rangle$ by
absorbing $L_1$ photons. On the other hand, since $K=\left\|\langle
1\vert H'_{\Lambda}\vert
C\rangle\right\|=\frac{1}{2}\sqrt{\Omega_S^2+\Omega_T^2}>0$, there
is nonzero probability for an atom in the coupled state $\vert
C\rangle$ to be excited to $\vert 1\rangle$. In contrast to the
excitation paths, the spontaneous decay processes for $\vert
1\rangle \rightarrow \vert NC\rangle$ and $\vert 1\rangle
\rightarrow \vert C\rangle$ do not contain such a clear-cut
asymmetry. The decay rates associated with $\vert 1\rangle
\rightarrow \vert C\rangle$ and $\vert 1\rangle \rightarrow \vert
NC\rangle$ are proportional to the square of their corresponding
dipole moments, i.e., $ \Gamma_{1\rightarrow
NC}=\left(\sin^2\chi\cos^2\theta
+\cos^2\chi\sin^2\theta\right)\Gamma_1$, and $\Gamma_{1\rightarrow
C}=\left(\sin^2\chi\sin^2\theta
+\cos^2\chi\cos^2\theta\right)\Gamma_1. $ Take $\chi=\theta=\pi/4$
as an example, for which $\Gamma_{1\rightarrow
NC}=\Gamma_{1\rightarrow C}=\Gamma_1/2$, it is clear that one
pump-decay cycle $\vert C\rangle \rightarrow \vert 1\rangle
\rightarrow \vert C \rangle \mbox{ or } \vert NC \rangle$ will
transfer half of the population of $\vert C\rangle$ into $\vert
NC\rangle$. After several pump-decay cycles, most atomic population
will be trapped in $\vert NC\rangle$, i.e., DSCPT is obtained.

It should be noted that $\vert NC\rangle$ itself is not radiatively
stable, due to the spontaneous decay from $\vert 2\rangle$ to $\vert
3\rangle$. To demonstrate this more explicitly we resort to the
fully-quantized correspondence of the semiclassical noncoupled state
$\vert NC\rangle$, i.e.,
\begin{equation}\label{eq.DSNC1}
\vert NC'_n\rangle =\cos\theta\vert S_n\rangle-\sin\theta\vert
T_n\rangle.
\end{equation}%
As shown in Ref.~\cite{Tannoudji}, $\vert S_n\rangle$ and $\vert
T_n\rangle$ spontaneously decay to the next low-lying doublet
composed of $\vert S_{n-1}\rangle$ and $\vert T_{n-1}\rangle$, by
emitting a reservoir photon whose frequency is centered at
$\omega_{23}$. Each spontaneous emission event (quantum jump)
completely destroys the coherence of $\vert NC'_n\rangle$, so that
the decoherence rate of $\vert NC'_n\rangle$ is equivalent to the
frequency of occurrence of such quantum jumps, i.e., the spontaneous
decay rate $\Gamma_2$ of state $\vert 2\rangle$. Nevertheless, if $K
\gg\Gamma_2$ and $\Gamma_{1\rightarrow NC}\gg\Gamma_2$, then after
each $\vert 2\rangle \rightarrow \vert 3\rangle$ quantum jump, the
DS coherence will be quickly reestablished in $\vert
NC'_{n-1}\rangle$ within a time scale much smaller than the average
time interval between two successive quantum jumps. In this sense
$\vert NC\rangle$ can be viewed as quasi-stable and we say that an
effective DSCPT occurs. Therefore, we expect to see fluorescence
suppression from $\vert 1\rangle$ and the population divergence
between $\vert NC\rangle$ and $\vert C\rangle$ at the vicinity of
two-photon resonance, as two symbolic manifestations for DSCPT.

Next we give a numerical calculation of DSCPT under the bare atomic
state basis, using a semiclassical master equation (see, e.g.,
\cite{Peng,HXM,ZYF}). In contrast to the theoretical analysis above,
the numerical method we use does not require a second RWA, so the
calculations are valid even when the driving fields are tuned far
away from the two-photon resonance condition (\ref{eq.2photon}).

%%%%%%%%%%%%%%%%%%%%%%%%%%%%%%%%%%%%%%%%%%%%%%%%%%%%%%%%%%%%%%%%%%%
%%%%%%%%%%%%%%%%%%%%%%%%%%%%%%%%%%%%%%%%%%%%%%%%%%%%%%%%%%%%%%%%%%%
\subsection{Numerical calculation}
For the model scheme shown in Fig.~\ref{fig1}(a) the semiclassical
Hamiltonian can be written as (with RWA)
\begin{equation}
H=\left[
\begin{array}{ccc}
\Delta_1 & 0 & \frac{1}{2}\Omega_1^+{\rm e}^{-{\rm i}\delta
t}+\frac{1}{2}\Omega_1^-{\rm e}^{{\rm i}\delta t} \cr 0 & \Delta_2 &
\frac{1}{2}\Omega_2 \cr \frac{1}{2}\Omega_1^+{\rm e}^{{\rm i}\delta
t}+\frac{1}{2}\Omega_1^-{\rm e}^{-{\rm i}\delta t} &
\frac{1}{2}\Omega_2 & 0
\end{array}\right].
\end{equation}%
The master equation governing the time evolution of atomic
populations and coherences can be obtained by combining the
reversible Liouville-equation ${\rm
i}\dot{\rho}=\left[H,\rho\right]$ with relevant irreversible
relaxation parameters, as,
\begin{equation}\label{mastereq}
\dot{x}(t)=\left[ A^-{\rm e}^{-{\rm i}\delta t}+A^0+A^+{\rm e}^{{\rm
i}\delta t}\right]x(t)+{\rm e}^{-{\rm i}\delta t}V^- +V^0+{\rm
e}^{{\rm i}\delta t}V^+.
\end{equation}%
In Eq.~(\ref{mastereq}),  $V^0$, $V^\pm$, and $x(t)$ are $8\times 1$
column vectors defined as , $ V^0=\left[\frac{1}{2}{\rm
i}\Omega_2,0,-\frac{1}{2}{\rm i}\Omega_2, 0,0,0,0,0\right]^T$, $
V^+=\left[0,\frac{1}{2}{\rm i}\Omega_1^-, 0,0,0,-\frac{1}{2} {\rm
i}\Omega_1^+,0,0\right]^T$, $ V^-=\left[0,\frac{1}{2}{\rm
i}\Omega_1^+, 0,0,0,-\frac{1}{2} {\rm i}\Omega_1^-,0,0\right]^T$ and
$x(t)=[x_1(t),\ldots,x_8(t)]^T$, where $x_1(t)=\rho_{23}$,
$x_2(t)=\rho_{13}$, $x_3(t)=\rho_{32}$, $x_4(t)=\rho_{22}$,
$x_5(t)=\rho_{12}$, $x_6(t)=\rho_{31}$, $x_7(t)=\rho_{21}$, and
$x_8(t)=\rho_{11}$. The notations $A^\pm$ and $A^0$ are $8\times 8$
matrices defined as $A^0_{11}={\rm i}\Delta_2-\gamma_{23}$,
$A^0_{22}={\rm i}\Delta_1-\gamma_{13}$, $A^0_{33}=-{\rm
i}\Delta_2-\gamma_{23}$, $A^0_{44}=-\Gamma_2$, $A^0_{55}={\rm
i}(\Delta_1-\Delta_2)-\gamma_{12}$, $A^0_{66}=-{\rm
i}\Delta_1-\gamma_{13}$, $A^0_{77}={\rm
i}(\Delta_2-\Delta_1)-\gamma_{12}$, $A^0_{88}=-\Gamma_1$,
$A^0_{43}=A^0_{76}=A^0_{67}=A^0_{38}=\frac{1}{2}{\rm i}\Omega_2$,
$A^0_{14}=-{\rm i}\Omega_2$,
$A^0_{41}=A^0_{52}=A^0_{25}=A^0_{18}=-\frac{1}{2}{\rm i}\Omega_2$,
$A^0_{34}={\rm i}\Omega_2$, $A^+_{71}=A^+_{82}=-\frac{1}{2}{\rm
i}\Omega_1^-$, $A^+_{24}=A^+_{17}=-\frac{1}{2}{\rm i}\Omega_1^+$,
$A^+_{64}=A^+_{35}=\frac{1}{2}{\rm i}\Omega_1^-$,
$A^+_{53}=A^+_{86}=\frac{1}{2}{\rm i}\Omega_1^+$, $A^+_{28}=-{\rm
i}\Omega_1^+$, $A^+_{68}={\rm i}\Omega_1^-$,
$A^-_{71}=A^-_{82}=-\frac{1}{2}{\rm i}\Omega_1^+$,
$A^-_{24}=A^-_{17}=-\frac{1}{2}{\rm i}\Omega_1^-$,
$A^-_{64}=A^-_{35}=\frac{1}{2}{\rm i}\Omega_1^+$,
$A^-_{53}=A^-_{86}=\frac{1}{2}{\rm i}\Omega_1^-$, $A^-_{28}=-{\rm
i}\Omega_1^-$, $A^-_{68}={\rm i}\Omega_1^+$, where
$\gamma_{12}=(\Gamma_1+\Gamma_2)/2$, $\gamma_{13}=\Gamma_1/2$ and
$\gamma_{23}=\Gamma_2/2$ are the damping rates of atomic coherences.
The remaining matrix elements of $A^\pm$ and $A^0$ are zeros. Note
that in deriving Eq.~(\ref{mastereq}) we have used
$\rho_{11}+\rho_{22}+\rho_{33}=1$ to eliminate $\rho_{33}$.

By decomposing $x(t)$ into a harmonic expansion
$x(t)=\sum_{n=-\infty}^{\infty}x^{(n)}(t){\rm e}^{{\rm i}\delta t}
$, Eq.(\ref{mastereq}) can be recast into an recursive relation
\begin{eqnarray}
\dot{x}^{(n)}(t)&=& \left[ A^0-{\rm i}n\delta I\right]x^{(n)}(t)+A^-x^{(n+1)}(t)\nonumber\\
&+&A^+x^{(n-1)}(t)+V^-\delta_{n,-1}+V^0\delta_{n,0}+V^+\delta_{n,1},
\end{eqnarray}%
where $I$ denotes a $8\times8$ unit matrix. In steady state,
$\dot{x}^{(n)}(t)=0$, which  gives the steady-state solution of
$x(t)$ below:
\begin{equation}\label{keyeq}
\left[\begin{array}{ccccc} \cdots\\ M_{-2}&-A^-\\
-A^+&M_{-1}&-A^-\\ &-A^+&M_{0}&-A^-\\ &&-A^+&M_{1}&-A^-\\
&&&-A^+&M_{2}\\ &&&& \cdots
\end{array}\right]
\left[\begin{array}{c}\cdots\\ x^{(-2)}(\infty)\\
x^{(-1)}(\infty)\\ x^{(0)}(\infty)\\ x^{(1)}(\infty)\\
x^{(2)}(\infty)\\ \cdots
\end{array}\right]
=\left[\begin{array}{c} \cdots\\ 0\\ V^- \\ V^0 \\ V^+\\ 0\\
\cdots\end{array}\right],
\end{equation}%
where $M_{n}={\rm i}n\delta I-A^0$. In this way we can calculate the
steady-state atomic density matrix by truncating Eq.~(\ref{keyeq})
at a given order with desired numerical precision. Moreover, using
Eq.~(\ref{eq.NC}) and (\ref{eq.C}), the steady-state population in
the $\vert NC\rangle$ and $\vert C\rangle$ can be calculated from
the solution of (\ref{keyeq}) straightforwardly. As an example for
$\chi=\theta=\pi/4$, the steady-state populations of $\vert
NC\rangle$ and $\vert C\rangle$ are
\begin{eqnarray}
\rho^{(0)}_{NC,NC}(\infty)&=&\frac{1}{2}
\left[1-x^{(0)}_8(\infty)+x^{(2)}_4(\infty)+x^{(-2)}_4(\infty)\right]\nonumber\\
&+&\frac{1}{4}\left[x^{(-2)}_8(\infty)-x^{(-2)}_3(\infty)+x^{(-2)}_1(\infty)\right]\nonumber\\
&+&\frac{1}{4}\left[ x^{(2)}_8(\infty)+ x^{(2)}_3(\infty)-
x^{(2)}_1(\infty)\right]
\end{eqnarray}%
and
\begin{equation}
\rho^{(0)}_{C,C}(\infty)=1-\rho^{(0)}_{NC,NC}(\infty)-x_8^{0}(\infty)
\end{equation}

%%%%%%%%%%%%%%%%%%%%%%%%%%%%%%%%%%%%%%%%%%%%%%%%%%%%%%%%%%%%%%%%%%%%
%Figure 2
\begin{figure}[htb]
\begin{center}
\includegraphics[bb=36 170 554 620, width=8cm, clip]{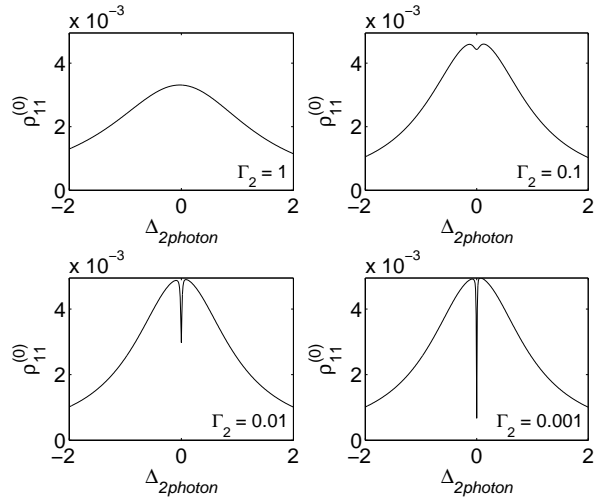}
\end{center}
\caption {The steady-state population of state $\vert 1\rangle$ as a
function of two-photon detuning $\Delta_{\rm 2photon}$. Here we
choose $\delta=5$, $\Delta_1=\Delta_2=0$, and $\Omega_1^+=\Omega_1^-
= 0.1$, so that $\chi=\theta=\pi/4$. The detuning $\Delta_{\rm
2photon}$ is realized via variation of $\Omega_2$, and all relevant
parameters are normalized with respect to $\Gamma_1$.}\label{fig2}
\end{figure}
%%%%%%%%%%%%%%%%%%%%%%%%%%%%%%%%%%%%%%%%%%%%%%%%%%%%%%%%%%%%%%%%%%%%
%Figure 3
\begin{figure}[htb]
\begin{center}
\includegraphics[bb=20 170 545 600, width=8cm, clip]{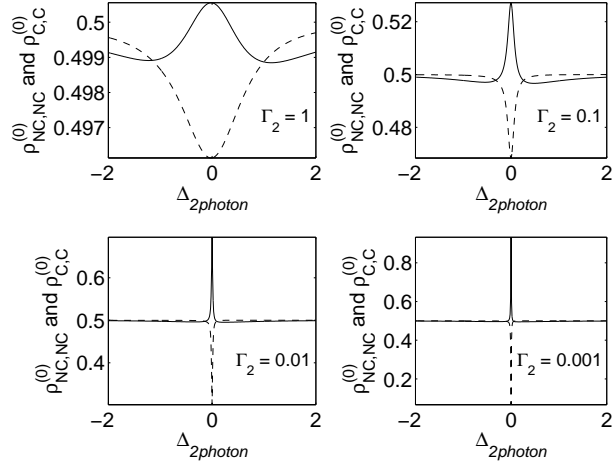}
\end{center}
\caption {The steady-state population in the noncoupled state $\vert
NC\rangle$ (solid line) and the coupled state $\vert C\rangle$
(dashed line) as a function of two-photon detuning $\Delta_{\rm
2photon}$. Here all parameters are the same as in
Fig.~\ref{fig2}.}\label{fig3}
\end{figure}

%%%%%%%%%%%%%%%%%%%%%%%%%%%%%%%%%%%%%%%%%%%%%%%%%%%%%%%%%%%%%%%%%%%
In real calculations, we use $\Gamma_1=1$ and normalize all
parameters with respect to $\Gamma_1$. Also, we specifically choose
$\Delta_2=0$ and $\Omega_1^+=\Omega_1^-$, to have
$\chi=\theta=\pi/4$ for simplicity. In Fig.~\ref{fig2} we show the
steady-state population $\rho_{11}^{(0)}=x_{8}^{(0)}(\infty)$ of
state $\vert 1\rangle$, as a function of two-photon detuning
$\Delta_{\rm 2photon}=\Omega_2-2\delta$ (since $\Delta_2=0$). Other
parameters are chosen as $\Delta_1=0$, $\Omega_1^+=\Omega_1^-=0.1$,
$\delta=5$, and $\Delta_{\rm 2photon}$ is varied through $\Omega_2$.
From Fig.~\ref{fig2} we see that, as $\Gamma_2$ decreases from 1 to
0.001, the darkline appears at $\Delta_{\rm 2photon}=0$ and
gradually becomes apparent. We also display in Fig.~\ref{fig3} the
steady-state populations of $\vert NC\rangle$ and $\vert C\rangle$,
using the same parameters as in Fig.~\ref{fig2}. For small
$\Gamma_2$, populations of $\vert NC\rangle$ and $\vert C\rangle$
demonstrate clear divergence at two-photon resonance, although they
are almost equally distributed at nonresonant places. The numerical
results also confirms that DSCPT is most effective for the parameter
range $K \gg \Gamma_2$ and $\Gamma_1 \gg \Gamma_2$, as demonstrated
in Fig.~\ref{fig2}(c) and Fig.~\ref{fig2}(d), as well as in
Fig.~\ref{fig3}(c) and Fig.~\ref{fig3}(d). We note that the widths
of the central dips and peaks in Fig.~\ref{fig2} and Fig.~\ref{fig3}
are in general slightly greater than $\Gamma_2$ due to the power
broadening introduced by $L_1$. As a summery of numerical
calculations we conclude that the parameter's range pertinent to
DSCPT can be given as $ \Omega_2 \gg K \gg \Gamma_2$, and $ \Gamma_1
\gg \Gamma_2$.

%%%%%%%%%%%%%%%%%%%%%%%%%%%%%%%%%%%%%%%%%%%%%%%%%%%%%%%%%%%%%%%%%%%
\section{Applications}\label{sec:3}
In this section we provide three examples on the application of
DSCPT. Throughout the discussions we refer a driven TLA to the
subsystem composed of transition $\vert2\rangle-\vert3\rangle$ and
field $L_2$. The DSCPT via inclusion of state $\vert1\rangle$ and
field $L_1$ is thus viewed as an artificial control over the driven
TLA. Moreover, all relevant parameters are normalized with respect
to $\Gamma_1$ as in the previous section.

\subsection{Precision measurement of Rabi frequency}
The absolute Rabi frequency is the product of the field strength and
the associated atomic dipole moment. Knowing either one, the
measurement of Rabi frequency reveals the other. Thus the precision
measurement of Rabi frequency is of great importance in laser and
atomic spectroscopy. The usual measurement of Rabi frequency invokes
the Autler-Townes (AT) spectroscopy, in which the Rabi splitting of
a resonantly driven TLA is probed by a second \textit{monochromatic}
field which couples the ground state $\vert3\rangle$ to a probe
state $\vert1\rangle$. The variation of the fluorescence intensity
from $\vert 1\rangle$ with respect to the detuning of the probe
field reveals an AT doublet. Spectroscopic measurement of the
frequency splitting of the AT doublet directly gives the Rabi
frequency $\Omega_2$.

The weakness of this conventional setup lies in that each absorption
peak of the AT doublet possesses an intrinsic linewidth (for, e.g.,
$\chi =\pi /4$)
\begin{equation}
W_{AT}= \frac{1}{2}\Gamma_2 +\Gamma_1,
\end{equation}%
which obviously depends on the spectral width of state $\vert
1\rangle$. This means that the introduction of a probe state
inevitably produces an additional measuring error $\Gamma_1$. As far
as monochromatic probe field is concerned, this error is unremovable
and the only way to reduce it is to choose a probe state with a
longer lifetime, which on the other hand leads to a longer photon
collection time and therefore deteriorates the signal-to-noise
ratio.

However, by using a \textit{bichromatic} probe field as described in
the previous discussion of DSCPT, the above difficulty can be
circumvented. The Rabi frequency $\Omega_2$ can be alternatively
measured through the internal frequency difference of the
bichromatic probe field, since the position of fluorescence darkline
(as shown in Fig.2) corresponds to the equality $\Omega_2=2\delta$.
Because the width of the dark resonance is independent of
$\Gamma_1$, $\Omega_2$ can be measured with a precision invulnerable
to the choice of the probe state. Although this precision is still
plagued by the $L_1$ field-induced power broadening, it can be made
sufficiently smaller than $\Gamma_1$ by reducing the intensity of
the bichromatic field $L_1$ to satisfy $K \ll \Gamma_1$. Therefore,
this method would be superior to the conventional AT spectroscopy
whenever the spectral width of the probe state is much larger than
that of the atomic transition to be measured.

\subsection{Unexpected population inversion}
It is well known that no steady-state population inversion exists
for a driven TLA~\cite{Loudon}. Sustained population inversion can
be realized in multi-level atoms where irreversible population
transfer channels between excited states come into play. For
example, if an additional $\vert 1\rangle \rightarrow \vert
2\rangle$ decay channel is open for the V-type atom shown in
Fig.~\ref{fig1}(a), a steady-state inversion on the $\vert 2\rangle
-\vert 3\rangle$ transition can be achieved with appropriate
parameter setup. However, even when no such decay channel is
available, artificial population transfer in the DS basis can still
lead to an unexpected inversion for the $\vert 2\rangle -\vert
3\rangle$ transition, which is termed as \textit
{dynamically-induced irreversibility} by Meduri \textit {et
al}~\cite{meduri}. In Section~\ref{sec:2} we have seen that with
ideal DSCPT ($\rho_{NC,NC}\approx1$) the atomic population is
trapped in the dark state $\vert NC\rangle$, thus the mixing angle
$\theta$ provides an independent doorknob on the DS populations.
Next we show that the unexpected inversion can be realized via DSCPT
as well, and the results in~\cite{meduri} can be viewed as a
limiting case of this method.

As shown in a rate-equation approach~\cite{Wilson}, the DS
populations of a monochromatically driven TLA obey
\begin{equation}\label{eq.dspop}
\frac{\rho_{SS}}{\rho_{TT}}=\tan^4\chi,
\end{equation}%
which leads to $\rho_{SS} \leqslant \rho_{TT}$ because $0 \leqslant
\chi \leqslant \pi/4$, i.e., the DS population is always
non-inverted. Furthermore, from Eq.(\ref{eq.ds.semi}) one finds
\begin{equation}\label{eq.inv}
\rho_{22}-\rho_{33}=\cos2\chi\cdot (\rho_{SS}-\rho_{TT}).
\end{equation}
Due to the nonnegativity of $\cos2\chi$, a DS inversion always
corresponds to a bare atomic inversion except at $\chi=\pi/4$. Such
connection between DS inversion and bare atomic inversion is the key
factor in achieving unexpected inversion.

When state $\vert1\rangle$ and field $L_1$ are included, equality
(\ref{eq.dspop}) can be broken and novel effects will occur. Under
ideal DSCPT configuration ($\rho_{NC,NC}\approx1$), the DS
populations is better approximated by
\begin{equation}\label{eq.approx}
\frac{\rho_{SS}}{\rho_{TT}}\approx\cot^2\theta,
\end{equation}
according to Eq.(\ref{eq.NC}). Thus a DS inversion can be achieved
for $\theta < \pi/4$ (i.e., ${\Omega_T}>{\Omega_S}$). Substitute
(\ref{eq.approx}) into (\ref{eq.inv}) (along with
$\rho_{SS}+\rho_{TT}$=1), we arrive at
\begin{equation}\label{eq.inv1}
\rho_{22}-\rho_{33}\approx\cos2\chi\cdot
\frac{\cot^2\theta-1}{\cot^2\theta+1},
\end{equation}
which is a concise prediction on the magnitude of bare atomic inversion
assuming ideal DSCPT.

%%%%%%%%%%%%%%%%%%%%%%%%%%%%%%%%%%%%%%%%%%%%%%%%%%%%%%%%%%%%%%%%%%%%
%Figure 4
\begin{figure}[htb]
\begin{center}
\includegraphics[bb=16 180 556 610, width=8cm, clip]{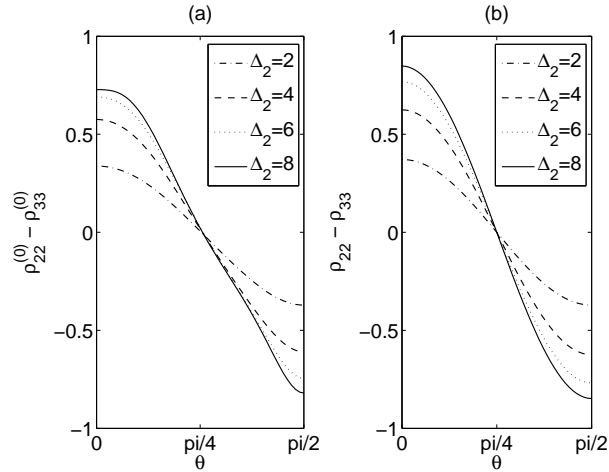}
\end{center}
\caption {The steady-state inversion $\rho_{22}-\rho_{33}$ as a
function of $\theta$ for several values of detuning $\Delta_2$,
which correspond to $\chi=$ 0.6 (dash-dotted), 0.45 (dashed), 0.34
(dotted) and 0.28 (solid). Plots in (a) and (b) are obtained by an
exact numerical calculation and an approximate analytical method,
respectively.}\label{fig4}
\end{figure}
%%%%%%%%%%%%%%%%%%%%%%%%%%%%%%%%%%%%%%%%%%%%%%%%%%%%%%%%%%%%%%%%%%%%

In Fig.\ref{fig4} we plot the steady-state inversion
$\rho_{22}-\rho_{33}$ as a function of $\theta$ numerically [(a),
using the procedure in Section~\ref{sec:2}] and analytically [(b),
using Eq.(\ref{eq.inv1})]. For a convenient comparison with existing
studies, we have chosen $\Gamma_2=0.0025$ to represent a real Barium
V-type atom spanned by $6s6p^1P_1$ ($\vert 1\rangle$), $6s6p^3P_1$
($\vert 2\rangle$) and $6s6s^1S_0$ ($\vert 3\rangle$). In all
calculations we fix $\Omega_2=5$ and tune $\chi$ by varying
$\Delta_2$. Within the range $0<\theta<\pi/2$
$\sqrt{\Omega_T^2+\Omega_S^2}=1$ is kept by adjust $\Omega_1^{\pm}$.
Meanwhile, the two-photon resonance condition for DSCPT is met by
choosing $\Delta_1=\frac{1}{2}\Delta_2$ and
$\delta=\frac{1}{2}\sqrt{\Delta_2^2+\Omega_2^2}$. As is seen from
Fig.\ref{fig4}, the numerical calculation matches very well with the
analytical approximation. The non-inverted to inverted transition
always occurs at $\theta=\pi/4$ as predicted from
Eq.(\ref{eq.inv1}). The highest inversion in Fig.\ref{fig4}(a) is
0.73, which is much larger than 0.11 reported in~\cite{meduri}, and
is very close to the best value 0.77 obtained by intense numerical
optimization in~\cite{HUPENG}.

In the limit ${\Omega_T}/{\Omega_S}\rightarrow \infty$
($\theta\rightarrow 0$) the bichromatic field $L_1$ reduces to a
monochromatic one. Under such condition our model will be
essentially the same as in~\cite{meduri}, except that the analysis
in Ref.~\cite{meduri} is based on first dressing the transition
$\vert 1\rangle -\vert 3\rangle$ instead of $\vert 2 \rangle -\vert
3\rangle$. Within their parameter range the magnitude of inversion
is relatively poor, as is verified by later numerical
simulations~\cite{HUPENG}. However, as shown by our analytical and
numerical investigation, the inversion via DSCPT provide not only a
better approaching to the optimal parameters but also a simpler
understanding on the origin of the unexpected inversion.

\subsection{LWI with and without hidden inversion}
As an attractive approach for the generation of continuous-wave
short-wavelength lasers, LWI in atomic systems has been well studied
for many years. Thank to the various quantum interference channels
available, multi-level systems are currently the workhorses for
LWI. However, two-level system is still a good demonstration of the
principles.

\begin{figure}[htb]
\begin{center}
\includegraphics[bb=30 180 550 590, width=8cm, clip]{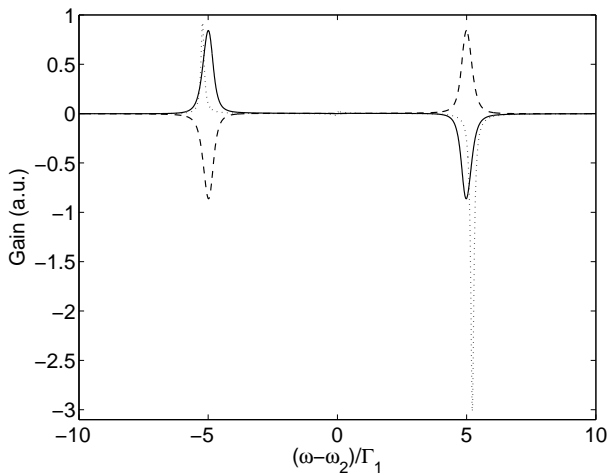}
\end{center}
\caption {LWI via hidden inversion. The dotted line denotes the
probe gain for a off-resonantly driven TLA, where an optimal
detuning ($\Delta_2=1.5$, $\Omega_2=5$, $\Gamma_2=0.1$) is used for
maximal gain~\cite{Wu}. LWI via DSCPT induced hidden inversion is
plotted for $\theta=0$ (dashed line) and $\theta=\pi/2$ (solid
line), with parameters $\Delta_2=0$, $\Gamma_1=1$, $\Gamma_2=0.1$,
$\Omega_2=5$, and $\sqrt{\Omega_T^2+\Omega_S^2}=0.5$.}\label{fig5}
\end{figure}

\begin{figure}[htb]
\begin{center}
\includegraphics[bb=20 180 550 610, width=8cm, clip]{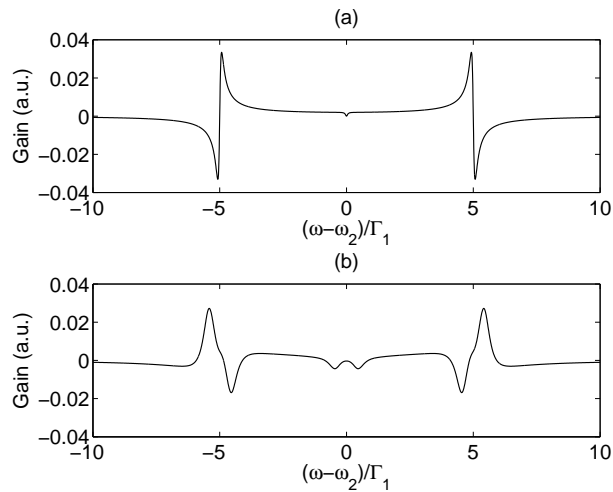}
\end{center}
\caption {LWI without hidden inversion. The probe gain in (a) a
resonantly driven TLA  ($\Delta_2=0$, $\Omega_2=5$, $\Gamma_2=0.1$)
and (b) a resonantly driven TLA with DSCPT ($\Omega_T=\Omega_S=1$,
$\Gamma_2=0.1$, $\Omega_2=5$, $\Delta_2=0$).}\label{fig6}
\end{figure}

Two LWI regimes are well recognized for a coherently driven TLA: (i)
gain via hidden inversion and (ii) gain without hidden inversion. In
the first regime, although bare atomic population is not inverted,
the net gain of a probe field can still be attributed to a
hidden inversion (e.g., inversion in the DS basis). A typical example is
the gain/absorption peak located at the left/right Rabi sideband for
a \textit {off-resonantly} driven TLA (see, the dotted line in
Fig.~\ref{fig5}), where the magnitude of gain/absortpion is proportional to the
DS population difference~\cite{Wilson}. In the second regime, the
probe gain originates from the coherent energy transfer between the
two fields and no population inversion is found for any meaningful
basis. An example is the probe gain in a \textit {resonantly}
driven TLA [see, Fig.~\ref{fig6}(a)]~\cite{Mollow}. Such gain
profile cannot be attributed to hidden inversion since the DS
population is now equalized. The gain in the second regime is
generally much weaker than that in the first regime.

Aided with DSCPT, an interesting LWI via hidden inversion can be
seen even when the driving field is on-resonance with the TLA. In
Fig.~\ref{fig5} we plot the probe gain for $\theta=0$ (dashed line)
and $\theta=\pi/2$ (solid line) respectively. In contrast to the LWI
for an off-resonantly driven TLA, where the gain is always much
weaker than the absorption, the probe gain/absorption peaks are now
equal in magnitude. To explain the probe spectra, it is convenient
to attribute the gain/absorption at the left (right) Rabi sideband
to the DS transition $\vert T_{n}\rangle \rightarrow \vert S_{n-1}
\rangle$/$\vert S_{n-1}\rangle \rightarrow \vert T_{n} \rangle$
($\vert S_{n}\rangle \rightarrow \vert T_{n-1} \rangle$/$\vert
T_{n-1}\rangle \rightarrow \vert S_{n} \rangle$). Recalling that the
total gain/absorption for each Rabi transition is proportional to
the corresponding DS population difference and the square of the
associated dipole moment [which is proportional to $\sin^4\chi$
($\cos^4\chi$) for the left (right) Rabi transition], the gain
profile in Fig.~\ref{fig5} can be well understood for $\chi=\pi/4$.
Another interesting feature takes place for $\theta=\pi/4$ (i.e.,
$\rho_{SS}=\rho_{TT}$), in this case no DS inversion exists at all.
The probe gain under such condition is plotted in
Fig.~\ref{fig6}(b). As expected, such LWI without hidden inversion
is much weaker than that shown via hidden inversion. In contrast to
Fig.~\ref{fig6}(a), the gain is now located at the \textit {outer}
sides rather than the \textit {inner} sides of the two Rabi
sidebands. Numerical calculations show (not shown here) a smooth
evolution between Fig.~\ref{fig6}(a) and (b) when $\Omega_{T}$ and
$\Omega_{S}$ are gradually increased. Such difference in gain
profile caused by increasing DS coherence is more or less connected
with the \textit{amplification by coherence} investigated
in~\cite{Agarwal}, however, to explore the detail of which is beyond
the scope of the present paper.
%%%%%%%%%%%%%%%%%%%%%%%%%%%%%%%%%%%%%%%%%%%%%%%%%%%%%%%%%%%%%%%%%%%%
%Figure 5
%%%%%%%%%%%%%%%%%%%%%%%%%%%%%%%%%%%%%%%%%%%%%%%%%%%%%%%%%%%%%%%%%%%%
%%%%%%%%%%%%%%%%%%%%%%%%%%%%%%%%%%%%%%%%%%%%%%%%%%%%%%%%%%%%%%%%%%%%
%Figure 6
%%%%%%%%%%%%%%%%%%%%%%%%%%%%%%%%%%%%%%%%%%%%%%%%%%%%%%%%%%%%%%%%%%%%
\section{Summary}\label{sec:4}
We have shown that by using an auxiliary bichromatic field, CPT can
be formed in the DS basis of a monochromatically driven TLA, using
both a dressed-atom analysis and a master equation calculation. The
demonstrated DSCPT can be viewed as a new doorknob to manipulate a
two-level system via its control over the DS populations. Examples
of such manipulation are discussed in various applications, among
which we show that the DSCPT induced unexpected inversion provides a
new point of view to the origin of dynamically induced
irreversibility, both qualitatively and quantitatively.

\ack

This work was supported by the PCSIRT, the SRFDP, the NFRPC Grant
No.~2006CB921205, and the National Natural Science Foundation of
China Grant No.~10625416 and No.~10534060.

\end{document}